\documentclass {imsart}

\usepackage{amsthm,amsmath,natbib}
\usepackage{graphicx,amsfonts,amssymb,mathabx}

\pdfoutput=1
\startlocaldefs

\newcommand{\nc}{\newcommand}
\newcommand{\R}{\mathbb R}

\newtheorem{thm}{Theorem}
\newtheorem{lemma}{Lemma}
\newtheorem{prop}[thm]{Proposition}

\nc{\Rd}{{\mathbb R}^d}

\endlocaldefs

\begin{document}

\begin{frontmatter}

\title{Distribution free testing for linear regression. Extension to general parametric regression}

\runtitle{Distribution free testing of linear regression}

\begin{aug}

\author{\fnms{Estate V.} \snm{Khmaladze} \thanksref{a} \ead[label=e1]{Estate.Khmaladze@vuw.ac.nz}}
\address[a]{Victoria University of Wellington, PO Box 600, Wellington, New Zealand \printead{e1}}

\runauthor{Estate Khmaladze}

\affiliation{Victoria University of Wellington}

\end{aug}

\begin{abstract}
Recently a distribution free approach for testing parametric hypotheses based on unitary transformations has been suggested in \cite{Khm13, Khm16, Khm17} and further studied in  \cite{Ngu17} and \cite{Rob19}. In this note we show that the transformation takes extremely simple form in distribution free testing of linear regression. Then we extend it to general parametric regression with vector-valued covariates.
\end{abstract}




\end{frontmatter}

 \section{Introduction. An illustrative example with linear regression}\label{intro}

The situation we consider in this paper is that of the classical parametric regression: given a sequence of  pairs of random variables $(X_i,Y_i)_{i=1}^n$, where $Y_i$ is  the response variable, while $X_i$ is the explanatory variable, or covariate, of this $Y_i$,  consider regression of $Y_i$ on $X_i$,
$$ Y_i = m(X_i) + \epsilon_i .$$
We assume that, given covariates $(X_j)_{j=1}^n$, the errors $(\epsilon_i)_{i=1}^n$ are i.i.d, and have expected value zero and finite variance -- for the sake of simplicity we assume this variance equal 1.

We are interested in the classical problem of testing that the regression function $m(x)$ belongs to a specified parametric family of functions $(m(x,\theta), \theta \in \Theta)$, which depend on a finite-dimensional parameter $\theta$ and which satisfy more or less usual regularity assumptions as functions of this $\theta$. 

Our aim is to describe a new method to build asymptotically distribution free theory for testing such hypothesis.  More specifically, we will construct asymptotically distribution free version of the regression empirical process, so that functionals from this process, used as test statistics, will be asymptotically distribution free.  The core of the method is based on the application of unitary operators as described more or less recently in \cite{Khm13, Khm16} and studied in \cite{Rob19} and \cite{Ngu17}. 

Earlier, asymptotically distribution free transformation of regression empirical process was suggested in \cite{KhmKou04}. For $d$-dimensional covariates, the limit distribution of the transformed process was that of standard Brownian motion on $[0,1]^d$. In this paper, the transformed process will converge to a standard projection of the standard Brownian motion on $[0,1]^d$, and the transformation will take surprisingly simple form, convenient in everyday practice. As  in \cite{KhmKou04}, this transformation is connected with no loss of statistical information.

The shortest way to show how the method works is to consider the most simple linear regression model. That is, in
\begin{equation}\label{lr0}
Y_i=X_i \theta + \epsilon _i, \; i=1,\dots, n, \; {\rm or\, in\, vector\, form,} \; \; \; Y = X\theta +\epsilon, \end{equation}
the covariates $X_i$, and the coefficient $\theta$ are one-dimensional. 
On probabilistic nature of the covariates $(X_i)_{i=1}^n$, we will make, practically, no assumptions. We only will use their empirical distribution function
$$F_n(x)=\frac{1}{n}\sum_{i=1}^n \mathbb{I}_{(X_i\leq x)} $$
and assume that as number of observed pairs $n$ increases it weakly converges to some limiting distribution $F$ -- an assumption of ergodic nature. Whenever we use time transformation $t=F(x)$, we will also assume that $F$ is continuous. All expectations below will be conditional expectations given the vector of numbers $(X_i)_{i=1}^n$. 

Consider estimated errors, or residuals,
$$ \hat \epsilon =Y - X \hat\theta \quad {\rm with} \;  \; \; \hat \theta = \langle Y,z\rangle ,$$
where $z=X/\langle X,X\rangle^{1/2}$ is the normalised vector of  covariates. The natural object to base a goodness of fit test upon is given by the partial sums process 
$$ \hat w_{n} (x) = \frac{1}{\sqrt{n}}\sum_{i=1}^n \hat \epsilon_i  \mathbb{I}_{(X_i\leq x)} .$$
However, the distribution of the vector $\hat \epsilon$ depends  on covariates: its covariance matrix has the form
$$E\hat \epsilon \, \hat \epsilon^T= I - z z^T .$$
As to the limit in distribution for the process $\hat w_{n} $, it is a projection of some 
Brownian motion, but not the Brownian bridge. Its distribution remains dependent on 
behaviour of the covariates. The limit distribution of statistics based on this process, and in particular, its supremum, will not be easy to  calculate.

However, consider new residuals obtained from $\hat \epsilon$ by unitary transformation 
$$
U_{a, b} = I - \frac{\langle a - b, \, \cdot  \,\rangle}{1-\langle a,  b \rangle}  (a - b) 
$$
with $n$-dimensional vectors $a$ and $b$ of unit norm: $\|a\|=\|b\|=1$. If $a=b$ we take $U_{a, b} = I$.  This operator in unitary, it maps $a$ into $b$ and $b$ into $a$, and it maps any vector $c$, orthogonal to $a$ and $b$, to itself, see, e.g., \cite{Khm13}, Sec. 2. Now choose $a=z$ and choose $b$ equal $r=(1,\dots,1)^T/\sqrt{n}$, the vector not depending on covariates at all.  Since the vector of residuals $\hat \epsilon$ is orthogonal to the vector $z$, we obtain:
$$\hat e= \hat \epsilon - \frac{\langle \hat \epsilon, r \rangle}{1-\langle z,r \rangle} (r-z) .$$
These new residuals have covariance matrix
$$E\hat e  \hat e^T= I - r r^T .$$
This would be the covariance matrix of the residuals in the problem of testing
\begin{equation}\label{shift}
Y_i=\theta + \epsilon_i , \quad i=1,2,\dots, n,\end{equation}
which is completely free from covariates. Yet, the transformation of $\hat \epsilon$ to $ \hat e$ is one-to-one
and therefore $\hat e$ contain the same ``statistical information", whichever way we measure it, as $\hat\epsilon$. One could say that the problem of testing linear regression \eqref{lr0} and testing \eqref{shift} is the {\it same} problem. 

The partial sum process based on the new covariates,
$$ \hat w_{n,e} (x) = \frac{1}{\sqrt{n}}\sum_{i=1}^n \hat e_i  \mathbb{I}_{(X_i\leq x)} ,$$
will converge in distribution, with time  transformation $t=F(x)$, to standard Brownian bridge. Therefore, limit distribution for all classical statistics will be free from covariates and known.

\begin{center}
\begin{figure}[h]
\includegraphics[scale=0.45]{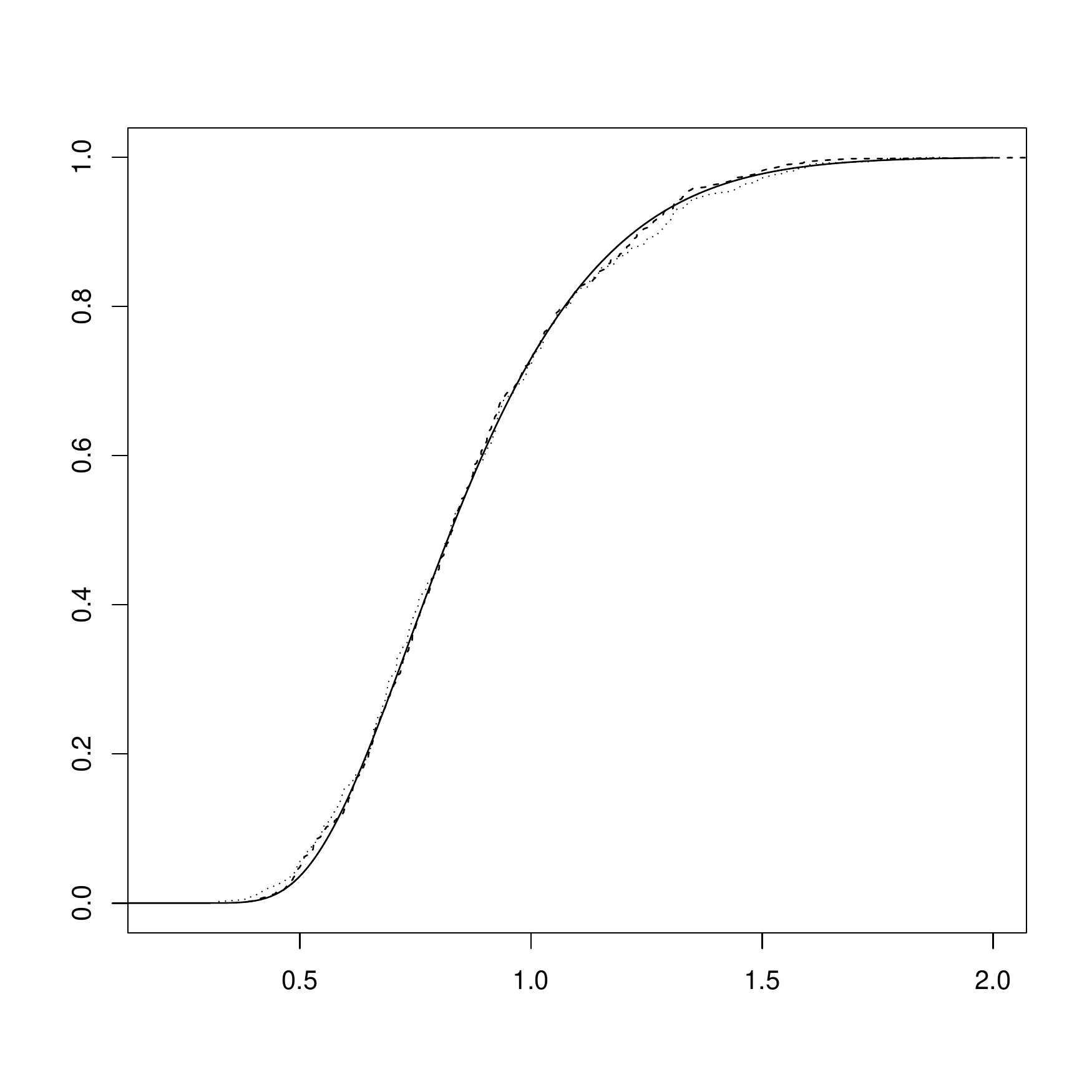}
\caption{The smooth line is Kolmogorov distribution function. The two other ones are simulated distributions of $\max _x|\hat w_{n,e} (x)|$ for two entirely different behaviour of covariates. In one case $X_i$-s have uniform distribution on $[0,2]$ while in the other they have Gaussian distribution $N(1,2)$. 200 replications of samples of size $n=200$.}
\end{figure}
\end{center}

Asymptotically distribution free tests, even if only for the case of linear regression, have been of main interest from long ago. To achieve this distribution free-ness different forms of residuals have been suggested, various decompositions of $z$, especially when covariates $X_i$ are multidimensional, have been studied and approximations for quadratic forms from $\hat \epsilon$ have been developed. Assumption of normality, arbitrary as it is in many cases, has been made more or less casually. If one is allowed somewhat free speech, one could say that a mathematical lace has been created. Good source for this material is the book \cite{CooWei82}. In dry residue. only the chi-square tests have been obtained. Distribution free forms of other classical statistics were never considered and constructed. We refer to \cite{McCNel08} for much of the existing theory for linear models. The most recent review on goodness of fit problems in regression which we know of is \cite{Gonz-Man_etal13}. 

Note that the initial regression process of this paper, not yet asymptotically distribution free, is different from what was used in previous work, including relatively recent ones.  Although partial sum processes, like $\hat w_n$, form one of the main objects of asymptotic theory, it is often that a different form of such processes is considered, one simple example of which would be
\begin{equation}\label{scan1}
\frac{1}{\sqrt{n}}\sum_{i=1}^n (X_i - \bar X_n)  \mathbb{I}_{(\hat \epsilon_i\leq x)} ,
\end{equation}
(see more sophisticated form of the weight function in recent paper \cite{ChoMue18}). Here the scanning over the values of the residuals is used. This is very natural way of scanning when the statistical problems considered pertain to distribution of errors. An example, studied in well known papers \cite{DetMun98}, \cite{DetHet09}, \cite{Detetal07} and loc.cit. \cite{ChoMue18}, is the problem of testing heterogeneity of errors. The same scanning is basically unavoidable in study of distribution of i.i.d. errors, cf. \cite{KouMueSch17}, and in analysis of the distribution of innovations in autoregression models, see \cite{MueSchWef09}.

In our current situation of testing the form of regression function, it is a natural wish to see, in the case there is a deviation from the model, for what region of values of the covariate the deviation takes place, and scanning in $X_i$-s will allow this. Even in the simple case when the covariate is just discrete time, taking values $1,2,\dots, n$, it would be strange not to examine the sequence $\hat \epsilon_1,\dots, \hat \epsilon_n$, in this time, but instead look on the order statistics based on them, which scanning as in \eqref{scan1} would imply. These considerations motivate the form of the regression process  $\hat w_n$ and $\hat w_{ne}$. To make the illustrative example of this section more of immediate practical use and to explain better the asymptotic behaviour of the regression empirical process, in the next Section  \ref{gen_lin} we consider the general form  of one-dimensional linear regression. In the following Section \ref{gen} we consider general parametric regression. In this case the time transformation, considered in (iii) of the Proposition \ref{prop2} below again leads to distribution free-ness if $F$ is continuous. If $F$ is discrete, then the method suggested in \cite{Khm13}, Sec. 2, can be easily used. In Section \ref{multi_time}  we consider multidimensional $X_i$s. Transformation fo $\hat \epsilon$ to $\hat e$ will not change, but to standardise distribution of regressors one could use normalisation by $\hat f_n^{1/2}$, where $\hat f_n$ is an estimator of the density of $F$, cf., e.g., \cite{EinKhm01},  
\cite{CanEinLae20}. Here, however, we consider an approach borrowed from the theory of optimal transportation, or Monge - Kantorovich transportation problem, see, e.g., \cite{Vil09}. Very interesting probabilistic/statistical  applications of this theory have been recently given in \cite{Hal18} and \cite{Seg18}.\\



\section{General linear regression on $\R$}\label{gen_lin}

Consider the standard linear regression on the real line,
\begin{equation}\label{lr}
Y_i=\theta_0 + X_i \theta_1 + \epsilon _i, \; i=1,\dots, n, \; \text{or\,} \; \; \; Y =\theta_0  {\bf 1} + X\theta_1 +\epsilon, \end{equation}
The $\bf 1$ here denotes a vector with all coordinates equal to the number $1$. Instead of \eqref{lr} consider its slightly modified and more convenient form
\begin{align}\label{lr1}
&Y_i=\theta_0 + (X_i  - \bar X) \theta_1 + \epsilon _i, \; i=1,\dots, n, \; \; 
\text{or\, in\, vector\, form,} \\
&Y =\theta_0  {\bf 1} + (X-\bar X {\bf 1}) \theta_1 +\epsilon, \notag
\end{align}
The least square estimations of $\theta_0$ and $\theta_1$ are
$$\hat\theta_0= \frac{1}{n}\sum_{j=1}^n Y_j \quad \text{and} \quad \hat \theta_1 = \frac{1}{\sum_{j=1}^n (X_j- \bar X)^2}\sum_{i=1}^n Y_j(X_j-\bar X) .$$
Using again notation $r$ and notation $$\tilde z= \frac{1}{\sqrt{\sum_{j=1}^n (X_j- \bar X)^2}} (X-\bar X),$$ for normalised vector of centered covariates, one can write the residuals as
$$
\hat\epsilon = Y - \hat \theta_0 {\bf 1} - \hat\theta_1 (X-\bar X {\bf 1}) 
$$
or in more succinct form
$$\hat\epsilon = Y - \langle Y,r\rangle r - \langle Y, \tilde z\rangle \tilde z .$$
Substitution of the linear regression model \eqref{lr1} for $Y$ produces representation of the vector of residuals $\hat \epsilon$ through the vector of errors $\epsilon$:
\begin{equation}\label{project0}
\hat\epsilon = \epsilon - \langle \epsilon, r\rangle r - \langle \epsilon, \tilde z\rangle\tilde z.
\end{equation}
This represents $\hat\epsilon$ as projection of $\epsilon$ orthogonal to $r$ and $\tilde z$.

From this it follows that the covariance matrix of $\hat\epsilon$ is
$$E\hat\epsilon\hat\epsilon^T = I - rr^T - \tilde z\tilde z^T,$$
and thus it still depends on the values of the covariates. The limit distribution of the regression process with these residuals,
$$ \hat w_{n} (x) = \frac{1}{\sqrt{n}}\sum_{i=1}^n \hat \epsilon_i  \mathbb{I}_{(X_i\leq x)} ,$$
will therefore have limit distribution which depends on $\tilde z$. 

It is possible to say more about the geometric structure of $ \hat w_{n} $ and its limiting process, and namely that
the limiting process will be a double projection of Brownian motion orthogonal to the functions $F(x)$ and
$$H(x)=\int^x h(y) dF(y), \; \text{ with} \; \; h(x) = \frac{x-\int y dF(y)} {\sqrt{\int (z-\int y dF(y))^2 dF(z)}}.$$
Here one can think of $h$ as a continuous time ``trace" of $\tilde z$.

To show this structure of  $ \hat w_{n} $ denote $\mathbb{I}_{x}$ the vector with 
coordinates $(\mathbb{I}_{(X_i\leq x)})_{i=1}^n$. Then we can write
$$\hat w_{n} (x) = \frac{1}{\sqrt{n}}\langle \hat\epsilon, \mathbb{I}_{x}\rangle = \frac{1}{\sqrt{n}}  \left [ \langle\epsilon, \mathbb{I}_{x}\rangle - \langle \epsilon, r\rangle \langle r , \mathbb{I}_{x} \rangle  - \langle \epsilon, \tilde z\rangle \langle \tilde z, \mathbb{I}_{x}\rangle \right] .$$
For the first term on the right hand side, considered as a process in $x$ and denoted $w_n(x)$, we can see that
\begin{equation}\label{BM}
w_n(x) = \frac{1}{\sqrt{n}} \langle\epsilon, \mathbb{I}_{x}\rangle = \frac{1}{\sqrt{n}} \sum_{i=1}^n \epsilon_i \mathbb{I}_{(X_i\leq x)}
\end{equation}
is the process of partial sums of i.i.d. random variables and $Ew_n^2(x) = F_n(x)$ while $F_n\to F$. Therefore, $w_n$ converges in distribution to Brownian motion in time $F$, i.e. $Ew_F^2(x) = F(x)$. Now consider the second term:  
$$ \frac{1}{\sqrt{n}} \langle \epsilon, r\rangle \langle r , \mathbb{I}_{x} \rangle = \frac{1}{\sqrt{n}} \sum_{j=1}^n \epsilon_j \, \frac{1}{n} \sum_{i=1}^n \mathbb{I}_{(X_i\leq x)} = w_n(\infty) F_n(x) .$$
The third term produces the following expression:
\begin{align*}
&\frac{1}{\sqrt {n}} \sum_{j=1}^n \epsilon_j (X_j-\bar X) \frac{1}{\sum_{j=1}^n (X_j- \bar X)^2}  \sum_{i=1}^n (X_i-\bar X) \mathbb{I}_{(X_i\leq x)} \\
&= \int (y- \bar X) w_n(dy) \frac{1}{\int (y-\bar X)^2 dF_n(y)} \int ^x (y-\bar X) dF_n(y)\\
&=  \int h_n(y) w_n(dy) \int ^x h_n(y) dF_n(y),
\end{align*}
where
$$ h_n(x)=\frac{x-\bar X}{\sqrt{\int (y-\bar X)^2 dF_n(y)}} .$$
This function, obviously, has unit $L_2(F_n)$-norm and is orthogonal to functions $const$ and $x$.
Overall, we see that 
\begin{equation}\label{project}
\hat w_{n} (x) = w_n(x) - w_n(\infty) F_n(x) - \int h_n(y) w_n(dy) \int ^x h_n(y) dF_n(y),
\end{equation}
and the right hand side of \eqref{project} is the orthogonal projector of $w_n$, which annihilates $F_n$ and $H_n$. As the consequence of this, if $\int y^2 dF(y)<\infty$, 
then $\hat w_\epsilon$ is the corresponding projection of the Brownian motion $w_F$.

What we propose now is, again, to replace the residuals $\hat\epsilon$ by another residuals, $\hat e$, constructed as their unitary transformation. As a preliminary step, assume that the covariates are listed in increasing order, $X_1<X_2<\dots<X_n$. One can assume this without loss of generality: even if it will entail re-shuffling of our initial pairs of observations, the probability measure we work under will not change, because the re-shuffled errors
will still be independent from permuted $(X_i)_{i=1}^n$ and will still form an i.i.d. sequence. 

Now introduce another vector $\tilde r$, different from $\tilde z$, which also has unit norm and is orthogonal to $r$. Define
$$\hat e = U_{\tilde z, \tilde r} \hat\epsilon = \hat \epsilon - \frac{\langle \hat \epsilon, \tilde r - \tilde z \rangle}{1-\langle z,r \rangle} (\tilde r-\tilde z) =  \hat \epsilon - \frac{\langle \hat \epsilon, \tilde r \rangle}{1-\langle \tilde z, \tilde r \rangle} (\tilde r-\tilde z) ,$$
where the second equality is true because the vector $\hat\epsilon$ is orthogonal to the vector $\tilde z$, see \eqref{project0}. Thus calculation of new residuals is as simple as in the previous case of \eqref{lr0}.

Let us summarise properties of $\hat e$ in the following proposition. In this, for transition to the limit when $n\to\infty$, it is natural to assume that $\tilde r_i$ can be represented through some piece-wise continuous function $\tilde r(t)$ on $[0,1]$:
\begin{equation}\label{newr}
\tilde r_i = \frac{1}{\sqrt n} \tilde r(\frac{i}{n}) ,
\end{equation}
in which case we have convergence
$$\frac{1}{\sqrt n} \sum_{i=1}^{nt} \tilde r_i = \frac{1}{n} \sum_{i=1}^{nt} \tilde r(\frac{i}{n}) \to \int_0^t \tilde r (s) ds = Q(t)
$$
and
$$
\sum_{i=1}^{nt} \tilde r_i^2 = \frac{1}{n} \sum_{i=1}^{nt} \tilde r^2(\frac{i}{n}) \to \int_0^t \tilde r^2(s) ds .
$$
Orthogonality of the vector $\tilde r$ to the vector $r$ implies orthogonality of the function $\tilde r(t)$ to functions equal constant, or $Q(1)=0$. For example, $\tilde r$ can be chosen as
\begin{equation}\label{newr1}
\tilde r_i = \sqrt{\frac{12}{n}}\left [\frac{i}{n} - \frac{n+1}{2n}\right] .
\end{equation}
\begin{prop} \label{prop1} (i) Covariance matrix of $\hat e$ is
$$E\hat e \hat e^T = I - r r^T - \tilde r \tilde r^T $$
and therefore does not incorporate covariates $X$ as soon as $\tilde r$ does not incorporate $X$.

(ii) If \eqref{newr} is true then the regression empirical process based on $\hat e$,
 $$\hat w_{n,e} (x)=\frac{1}{\sqrt{n}}\sum_{i=1}^n \hat e_i  \mathbb{I}_{(X_i\leq x)} $$
has the covariance function
$$E\hat w_{n,e} (x) \hat w_{n,e} (y) = F_n(\min(x,y)) - F_n(x)F_n(y) - Q_n(F_n(x)) Q_n(F_n(y)) + O(1/n),$$
where $Q_n (t) = \sum_{i=1}^{nt} \tilde r(\frac{i}{n})/n$.
In the case of \eqref{newr1} $$Q(F_n(x)) \sim - \sqrt{3}F_n(x) (1 - F_n(x)), \; n\to\infty .$$ 

(iii) As a corollary of (ii), the process $\hat w_{n,e} $, with change of time $t=F(x)$, converges in distribution to projection of standard Brownian motion on $[0,1]$ orthogonal to functions $1$ and $\tilde r$.
\end{prop}
The main step in the proof of $(i)$ is to express $\hat e$ through $\epsilon$:
\begin{align*}
U_{\tilde z, \tilde r} \hat\epsilon  &= U_{\tilde z, \tilde r} \epsilon - \langle \epsilon, r\rangle U_{\tilde z, \tilde r} r - \langle \epsilon, \tilde z\rangle U_{\tilde z, \tilde r} \tilde z\\
&= U_{\tilde z, \tilde r} \epsilon - \langle \epsilon, r\rangle r - \langle \epsilon, \tilde z\rangle \tilde r,
\end{align*}
where the second equality is correct because $r\perp \tilde z,\tilde r$ and $U_{\tilde z, \tilde r} \tilde z = \tilde r$ by the definition of $U_{\tilde z, \tilde r}$. Therefore
$$
\hat e = U_{\tilde z, \tilde r} \hat\epsilon = \epsilon - \frac{\langle \epsilon, \tilde r \rangle}{1-\langle \tilde z, \tilde r \rangle} (\tilde r-\tilde z) - \langle \epsilon, r\rangle r - \langle \epsilon, \tilde z\rangle \tilde r .$$
Calculation of the covariance matrix of the right hand side is now not difficult using shorthand formulas $E\epsilon \langle \epsilon, a \rangle = a$ and $E\langle\epsilon, a\rangle \langle \epsilon, b \rangle = \langle a, b \rangle$. After some algebra we obtain the expression given in (i). 

To show (ii) use vector notation for  $\hat w_{n,e} $:
$$E\hat w_{n,e} (x) \hat w_{n,e} (y) =\frac{1}{n}E \langle\mathbb{I}_{x},\hat e\rangle \langle \hat e, \mathbb{I}_{y} \rangle = \frac{1}{n}\mathbb{I}_{x}^T (I - r r^T - \tilde r \tilde r^T ) \mathbb{I}_{y} $$
Opening the brackets in the last expression one can find that
$$\frac{1}{n}\langle\mathbb{I}_{x}, \mathbb{I}_{y} \rangle = F_n(\min(x,y)) \quad \text{and} \quad  \frac{1}{n}\langle\mathbb{I}_{x}, r \rangle \langle\mathbb{I}_{y}, r \rangle =  F_n(x) F_n(y) ,$$
while
\begin{align*}
\frac{1}{n}\langle\mathbb{I}_{x}, \tilde r \rangle \langle\mathbb{I}_{y}, \tilde r \rangle &= \frac{1}{n} \sum_{i=1}^n \tilde r(\frac{i}{n}) \mathbb{I}_{(X_i\leq x)} \frac{1}{n} \sum_{i=1}^n \tilde r(\frac{i}{n}) \mathbb{I}_{(X_i\leq y)} \\
&= \frac{1}{n} \sum_{i=1}^{n F_n(x)} \tilde r(\frac{i}{n}) \; \frac{1}{n} \sum_{i=1}^{nF_n(y)} \tilde r(\frac{i}{n}) = Q_n(F_n(x))  Q_n(F_n(y))   \end{align*}
which proves (ii). 

The statement (iii) follows if we note that the covariance function of $\hat w_{n,e} (x)$ in time $t=F(x)$ converges to $\min(t,s) -ts-Q(t)Q(s)$, and that orthogonality of function $\tilde r(\cdot)$ to the function identically equal 1 makes the last expression the covariance of the Gaussian process $$w(t) - tw(1) - Q(t)\int_0^1 \tilde r(s) w(ds) ,$$ which indeed is the projection described in (iii). \hfill $\square$\\

In both regression models \eqref{lr0} and  \eqref{lr1} the process $\hat w_{n}$ turns out to be a projection of a Brownian motion, but for different values of covariates these projections are different. However, it is geometrically clear that it should be possible to rotate one projection into another, and this another into still another one, thus creating a class of equivalent projections -- those which can be mapped into each other. Then one can choose a single representative in each equivalence class, call it standard, and rotate any other projection into this standard one. What was done in this and the previous section was that we selected two standard projections and constructed the rotation of the other ones into these two.

The usefulness of this approach depends on how practically simple the rotation will be. For us, the transformations of $\hat\epsilon$ into $\hat e$ looks very simple. 

Finally, note that the model \eqref{lr1} includes two estimated parameters while the model \eqref{lr0} -- only one. However, since the vector $r$ is already ``standard", independent from covariates, there is no need to ``rotate" it to any other vector. Therefore in both cases one-dimensional rotation is sufficient. Situation when one needs to rotate several vectors at once, as well as general form of parametric regression, will be considered in the next Section \ref{gen}. \\

\section{General parametric regression}\label{gen}

Now consider testing regression model
\begin{equation}\label{gl}
Y_i=m_\theta(X_i) + \epsilon_i, \; i=1,\dots,,n,\;\text{ or in vector form,} \; Y=m_\theta(X) + \epsilon,
\end{equation}
where $m_\theta(X)$ denotes a vector with coordinates $(m_\theta(X_i))_{i=1}^n$, and $m_\theta$ is 
regression function, depending on $d$-dimensional parameter $\theta$. We will assume some regularity of 
$m_\theta(X_i)$ with respect to $\theta$, namely that $m_\theta(X_i)$  is continuously  differentiable in $\theta$. Obvious example when this condition is true is given by polynomial regression
$$m_\theta(x)=\theta_1 p_1(x) + \theta_2 p_2(x)+\dots+\theta_d p_d(x) ,$$
where $p_j(x), j=1,\dots, d,$ may form a system of (orthogonal) polynomials, or splines (see, e.g., \cite{Har15}, Sec.2.4.3), or trigonometric polynomials. There certainly are also many examples where $m_\theta(x)$ is not linear in $\theta$. 

Now denote
$$\dot m_\theta(x) = (\frac{\partial}{\partial\theta_1}m_\theta(x),  \dots, \frac{\partial}{\partial\theta_d}m_\theta(x))^T$$ 
a $d$-dimensional vector-function of the partial derivatives. Then $(\dot m_\theta(X_i) )_{i=1}^n$ is 
$d\times n$-matrix, with $d$ rows and $n$ columns. We assume that for every $\theta$ coordinates of $\dot 
m_\theta(x)$ are linearly independent as functions of $x$, which heuristically means that the model does not include unnecessary parameters.

Let now $\hat\theta$ denote the least square estimator of $\theta$, which is an appropriate solution of the  least squares' equation
$$
\sum_{i=1}^n \dot m_{\hat\theta}(X_i) \left [Y_i - m_{\hat\theta}(X_i) \right ] = 0 .
$$
Without digressing to exact justification (which can be found, e.g., in \cite{BatWat07}) assume that Taylor expansion in $\theta$ is valid and that together with normalization by $\sqrt{n}$ it leads to
$$
\frac{1}{\sqrt{n}} \sum_{i=1}^n \dot m_{\theta}(X_i) \left [Y_i - m_{\theta}(X_i) \right ] - R_n \sqrt{n} (\hat \theta - \theta) + \rho_n=0
$$
with a non-degenerate  $d\times d$-matrix $R_n$,
$$ R_n=\frac{1}{n}  \sum_{i=1}^n \dot m_{\theta}(X_i) \dot m_{\theta}^T(X_i) = \int \dot m_{\theta}(x)\dot m_{\theta}^T(x) dF_n(x) ,$$
and  $d$-dimensional vector of residuals $\rho_n$, such that $E\|\rho_n\|^2\to 0, n\to\infty$. Below for the terms asymptotically negligible in this sense we will use notation $o_P(1)$. From the previous display we obtain asymptotic representation for $\hat \theta$:
$$
\sqrt{n} (\hat \theta - \theta)=R_n^{-1} \frac{1}{\sqrt{n}} \sum_{i=1}^n \dot m_{\theta}(X_i) \left [Y_i - m_{\theta}(X_i)\right ] + o_P(1).
$$
As the final step, expand the differences $Y_i-m_{\hat \theta}(X_i)$ in $\theta$  up to linear term and substitute the expression for  $\sqrt{n} (\hat \theta - \theta)$ to get
$$
Y_i-m_{\hat \theta}(X_i) = Y_i-m_{\theta}(X_i) - \dot m_{\theta}^T (X_i) R_n^{-1} \frac{1}{n} \sum_{j=1}^n \dot m_{\theta}(X_j) [Y_j - m_{\theta}(X_j) ]  + o_P(1)
$$
or
$$
\hat \epsilon_i = \epsilon_i - \dot m_{\theta}^T (X_i) R_n^{-1} \frac{1}{n} \sum_{j=1}^n \dot m_{\theta}
(X_j) \epsilon_j +o_P(1) .
$$
In vector form this becomes
\begin{equation}\label{project1} 
\hat \epsilon = \epsilon - \dot m_{\theta}^T R_n^{-1} \frac{1}{n} \langle \dot 
m_{\theta}, \epsilon \rangle + o_P(1) , 
\end{equation}
an expression directly analogous to \eqref{project0}. It also describes the vector of residuals as being, asymptotically, projection of the vector of errors $\epsilon$, parallel to $d$ $n$-dimensional vectors of derivatives 
\begin{equation*}
(\frac{\partial}{\partial\theta_1}m_\theta(X_i))_{i=1}^n,  \dots, (\frac{\partial}{\partial\theta_d}m_\theta(X_i))_{i=1}^n .
\end{equation*}

It will be notationally simpler, while computationally not difficult, to change these linearly independent vectors to orthonormal vectors. Namely, introduce the functions
$$\mu_{\theta k} (x) = R_n^{-1/2} \frac{\partial}{\partial\theta_k}m_\theta(x), \quad k=1,\dots, d ,$$
and then the vectors
\begin{equation}\label{newmu}
\mu_{\theta k,i}= \frac{1}{\sqrt n} \mu_{\theta k} (X_i) , \; i=1,\dots, n.
\end{equation}
The two notations are convenient each in its place: $\mu_{\theta k}$ as a vector in $\R^n$  will be useful in expressions like \eqref{project2}, and $\mu_{\theta k} (\cdot)$ as a function in $L_2(F_n)$ will be useful in integral expressions like \eqref{project3}. Their respective norms are equal:
$$\sum_{i=1}^n \mu_{\theta k, i}^2 =\int \mu_{\theta k}^2(x) dF_n(x) .$$
 Which of these two objects we use will be visible in notation and clear from the context.
 
Now we can write \eqref{project1} as
\begin{equation}\label{project2}
\hat \epsilon = \epsilon - \sum_{k=1}^d \mu_{\theta k}^T  \langle \mu_{\theta k} , \epsilon \rangle + o_P(1) , 
\end{equation}
where the leading term on the right hand side is the projection of $\epsilon$ orthogonal to vectors 
$\mu_{\theta k}$. As a consequence, one can show that the following analogue of the representation \eqref{project} is true:
\begin{align}\label{project3}
\hat w_n(x)&=\frac{1}{\sqrt{n}}\sum_{i=1}^n [Y_i-m_{\hat \theta}(X_i)]{\mathbb I}_{(X_i\leq x)} \notag  \\ &= w_n(x) - \sum_{k=1}^d \int_{z\leq x} \mu_{\theta k} (z) dF_n(z) \int \mu_{\theta k} (z) w_n(dz) +o_P(1).
\end{align}
This, again, describes $\hat w_n$ as asymptotically projection of $w_n$ orthogonal to the functions $(\mu_{\theta k})_{k=1}^d$. We are ready to describe rotation of this projection to another, standard, projection, and of $\hat \epsilon$ to a vector of another residuals.

With some freedom of speech, we say that one can choose these new residuals in any way we wish; for example, choose them independent of any covariates. In particular, let $r_1(\cdot)$ be a function on $[0,1]$, identically equal $1$, and with this let vectors $r_k$ be defined as $r_{ki}=r_k(i/n)/\sqrt{n}$, where the system of functions $(r_k(\cdot))_{k=1}^d$ is such that
$$ \frac{1}{n}\sum_{i=1}^n r_k(\frac{i}{n}) r_l(\frac{i}{n}) =  \delta_{k,l}, \; k,l=1,\dots, d.$$
If we derive a unitary operator $K$,  which maps orthonormal vectors $(\mu_{\theta k})_{k=1}^d$ into vectors $(r_k)_{k=1}^{d}$, then this operator will map $\hat \epsilon$ into $\hat e$, and the covariance matrix of these new residuals will be defined solely by $(r_k)_{k=1}^d$ or $(r_k(\cdot))_{k=1}^{d}$.

As a side and rather inconsequential remark we note that it would be immediate to choose orthonormal polynomials on $[0,1]$, i.e. such that $$\int_0^1 r_k(s) r_l(s) ds=\delta_{k,l} ,$$  
which are continuous and bounded functions. Such polynomials will not satisfy the orthogonality condition in the previous display, but will require small corrections, asymptotically negligible for $n\to\infty$. If we insert these corrections in our notation it will make the text more complicated without opening any new feature of the transformation we want to discuss. Therefore in notations we will identify orthogonal polynomials in continuous time with those, orthonormal on the grid $\{1/n,2/n,\dots,1\}$.

It is essential that the structure of $K$ allows convenient handling. We present it here as a product of one-dimensional unitary operators. This allows coding of $K$ in a loop, and was tried for the case of contingency tables with about 30-dimensional parameter in \cite{Ngu17}. 

Suppose in one-dimensional unitary operator $U_{a,b}$ we choose $a=\mu_{\theta 1}$ and $b=r_1$ and apply the resulting operator $U_{\mu_{\theta 1}, r_1}$ to vector $r_2$:
$$U_{\mu_{\theta 1}, r_1}r_2=\tilde r_2 .$$
Then the product 
$$K_2=U_{\mu_{\theta 2}, \tilde r_2} \times U_{\mu_{\theta 1}, r_1}$$
is unitary operator which maps vectors $r_1, r_2$ to vectors $\mu_{\theta 1}, \mu_{\theta 2}$ and vice versa, and leaves vectors orthogonal to these four vectors unchanged. For a general $k$, define $\tilde r_k$ as
$$K_{k-1} r_k=\tilde r_k , \quad k=2,\dots, d.$$
\begin{lemma}\label{lem:rotation}
The product
\begin{equation*}
K_d=U_{\mu_{\theta d}, \tilde r_d}\times\dots\times U_{\mu_{\theta 1}, r_1}
\end{equation*}
is the unitary operator which maps $(r_k)_{k=1}^d$ to $(\mu_{\theta k})_{k=1}^d$ and vice versa, and leaves vectors orthogonal to $(r_k)_{k=1}^d$ and $(\mu_{\theta k})_{k=1}^d$ unchanged. 
\end{lemma}
The proof of this lemma was given, e.g., in \cite{Khm16}, section 3.4. It may be of independent interest for statistics of directional data, when explicit expression for rotations is needed. Therefore, for reader's convenience, at the end of this section we give  an essentially shorter proof. 

Thus, in proposition below we denote
\begin{equation}\label{param-e}
\hat e = K_d \hat \epsilon,
\end{equation}
and recall that $X_i$-s are numbered in increasing order. We also say 
$$E\hat \epsilon \hat \epsilon^T \sim I - \sum_{k=1}^d  \mu_{\theta k} \mu_{\theta k}^T$$
in the sense that for any sequence of $n$-vectors $b_n$, such that $\langle b_n, b_n\rangle \to c<\infty$
$$ E\langle b_n,\epsilon\rangle^2 \sim \langle   b_n, b_n\rangle  - \sum_{k=1}^d  \langle b_n, \mu_{\theta k}\rangle ^2, \; n\to \infty.$$ 
This notion of equivalence is used in the proposition below.
\begin{prop}\label{prop2} Suppose the regression function $m_\theta(x)$ is regular, in the sense that, for every $\theta$, the matrix $R_n$ is of full rank and converges to a matrix $R$ of full rank, and \eqref{project2} is true. Suppose the functions $r_k(\cdot), k=1,\dots,d,$ are continuous and bounded on $[0,1]$.Then

(i) for the covariance matrix of residuals $\hat e$ the following is true: 
\begin{equation*}
E\hat e \hat e^T \sim I - \sum_{j=1}^d  r_k r_k^T, \; n\to\infty;
\end{equation*}

(ii) for the empirical regression process, based on residuals $\hat e$ of \eqref{param-e},
$$\hat w_{n,e}(x) = \frac{1}{\sqrt {n}}\sum_{i=1}^n \hat e_i {\mathbb{I}}_{(X_i\leq x)},$$
the following convergence of the covariance function is true:
$$E\hat w_{n,e}(x) \hat w_{n,e}(y) \to F(\min(x,y)) - \sum_{j=1}^d Q_k(F(x)) Q_k(F(y)) , \text {as} \; n\to \infty ,$$
where $Q_k(t)=\int_0^t r_(s) ds$;\\
moreover,\\
(iii) the process $\hat w_{n,e}$, with time change $t=F(x)$ converges in distribution to projection of standard Brownian motion on $[0,1]$ orthogonal to functions $r_j(\cdot), j=1,\dots, d$.
\end{prop}

To prove (i) we do not need the explicit form of the operator $K_d$, and instead note that according to \eqref{project2}, up to asymptotically negligible term, $\hat \epsilon$ is projection of $\epsilon$, orthogonal to collection of $n$-vectors $\mu_{\theta 1},\dots,\mu_{\theta d}$. According to the lemma above, these vectors are mapped by operator $K_d$ to $n$-vectors $r_1,\dots, r_d$, and the operator $K_d$ is unitary. Therefore the vector $\hat \epsilon$ will be mapped into the vector which, up to asymptotically negligible term, will be projection of $\epsilon$ orthogonal to  $r_1,\dots, r_d$:
\begin{equation}\label{strong_map}
\hat e = \epsilon - \sum_{k=1}^d r_k \langle r_k,\epsilon\rangle + o_P(1) .
\end{equation} 
And the covariance matrix of this vector is the expression given in (i).

To prove (ii), replace $\hat e$ by its main term in \eqref{strong_map} in the expected value 
$$E\hat w_{n,e}(x) \hat w_{n,e}(y) =\frac{1}{n}E \langle\mathbb{I}_{x},\hat e\rangle \langle \hat e, \mathbb{I}_{y} \rangle \sim \frac{1}{n}\mathbb{I}_{x}^T (I - \sum_{k=1}^d r_k r_k^T) \mathbb{I}_{y}.$$
Here, since every $r_k(\cdot)$ is continuous and bounded,
$$\frac{1}{\sqrt n}\mathbb{I}_{x}^T  r_k = \frac{1}{n} \sum_{i=1}^n  r_k(\frac{i}{n}) \mathbb{I}_{(X_i\leq x)} \sim \int_{z\leq x} r_k(F(z)) dF(z) .$$

Statement (iii) of convergence in distribution follows not from unitarity property of $K_d$ as such, but from simplicity of its structure, reflected by \eqref{strong_map}. We have
\begin{equation*}
\hat w_{n,e}(x) \sim \frac{1}{\sqrt n} \langle\mathbb{I}_{x},\epsilon - \sum_{j=1}^d r_j \langle r_j, \epsilon\rangle \rangle = \frac{1}{\sqrt n} \langle\mathbb{I}_{x},\epsilon \rangle - \frac{1}{\sqrt n} \sum_{k=1}^d \langle\mathbb{I}_{x},r_k \rangle  \langle r_k, \epsilon\rangle
\end{equation*}
The first inner product on the right side, denoted $w_n(x)$ in \eqref{BM}, converges in distribution to $F$-Brownian motion. Expression for $\langle\mathbb{I}_{x},r_k \rangle $ we considered above, while
$$\langle r_j, \epsilon\rangle = \frac{1}{\sqrt n}\sum_{i=1}^n  r_k(\frac{i}{n}) \epsilon_i = \frac{1}{\sqrt n}\sum_{i=1}^n  r_k(F_n(X_i)) \epsilon_i= \int r_k(F_n(x)) w_n(dx) .$$
Thus, overall representation of $\hat w_{n,e}$ through $w_{n} $ has the form
\begin{equation}\label{estBM}\hat w_{n,e}(x) \sim w_{n}(x) - \sum_{k-1}^d \int_{z\leq x} r_k(F_n(z)) dF_n(z) \int r_k(F_n(x)) w_n(dx) . 
\end{equation}
Since $w_n$ converges in distribution to the $F$-Brownian motion $w_F$, which in time $t=F(x)$ becomes a standard Brownian motion $w$ on $[0,1]$, we see that the process $\hat w_{n,e}$ converges in distribution to the Gaussian process given by the right hand side of \eqref{estBM}, which in time $t=F(x)$ can be written as
$$\hat w (t) = w(t) - \sum_{k=1}^d Q_k(t) \int r_k(s) w(ds).$$
This is an orthogonal projection of $w$ orthogonal to the functions $r_j(\cdot), j=1,\dots, d$.\hfill $\square$

{\bf Proof of Lemma \ref{lem:rotation}.} Suppose $K_{k-1} r_j=\mu_{\theta j}, 1\leq j \leq k-1$; then it follows that $\tilde r_k\perp \mu_{\theta j}$, because $r_k \perp r_j$, and operator $K_{k-1}$ is unitary. But then , by its construction, $K_k r_j = U_{\mu_{\theta k}, \tilde r_k} K_{k-1} r_j=U_{\mu_{\theta k}, \tilde r_k}\mu_{\theta j} = \mu_{\theta j},$ while $K_k r_k= U_{\mu_{\theta k}, \tilde r_k} \tilde r_k = \mu_{\theta k}.$ Then the rest follows by induction.  \hfill $\square$

\section{The case of multi-dimensional covariates}\label{multi_time}

It is an important case when the covariate is a finite-dimensional vector. Let us use $p$ for dimension of each $X_i$. Again, we will not assume anything about probabilistic nature of these covariates, except that 
$$F_n(x)=\frac{1}{n}\sum_{i=1}^n {\mathbb I}_{\{X_i\leq x\}} \to F(x), $$
where $F$ is an absolutely continuous distribution function in $\R^p$. For simplicity of presentation, it will be convenient, however, to assume that $F$ is replaced by its copula function, or, equivalently, $F$ itself is supported on $[0,1]^p$, although the support can be a proper subset of $[0,1]^2$.

For $p$-dimensional  time, we could have shown that \eqref{project3} in the previous section is still correct. 
One of the relatively familiar ways to obtain distribution-free transformation of this process would be then to use the scanning martingale's approach of \cite{KhmKou04} to the projection \eqref{project3}. Another possibility would be to use unitary transformations suggested in \cite{Khm16} to map the projection \eqref{project3}  into another ``standard" projection, changing simultaneously the functions $\mu_{\theta k}(\cdot)$ and distribution $F$ to the corresponding objects of our choice. In doing this one will need to use estimator of the density of $F$. Here, however, we will see that both tasks can be achieved, again simultaneously but simpler, using the approach suggested by the theory of optimal transport.

For distribution free-ness of the vector of new residuals it does not matter how do we realise the  vectors $(r_k)_{k=1}^d$. For example, one can represent them in literary the same way as in \eqref{newr} -- the covariance matrix of the new residuals will depend on $r(\cdot)$ and not on covariates. However, similarly to \eqref{newmu}, see also discussion following \eqref{project3}, it will be very natural to connect  vectors $(r_k)_{k=1}^d$ with a system of piecewise continuous orthogonal functions $r_k(\cdot)$ of $p$ variables. 
To do this let us generate an i.i.d. sequence $(\xi_i)_{i=1}^n$ of random variables uniformly distributed on $[0,1]^p$. One could speak here about some distribution $G$ instead of the uniform distribution, but it will be a trite generality. The random variables $(\xi_i)_{i=1}^n$ will not be used to randomise our procedure but to serve as an ``anchor" to connect covariates $(X_i)_{i=1}^n$ to new ones which are uniformly distributed on $[0,1]^p$. 

Consider a one-to-one map $T$ of $(X_i)_{i=1}^n$ to $(\xi_i)_{i=1}^n$, so that $T(X_i)=\xi_j$ for one and only one $j$, cf. \cite{PeyCut19}, Sec. 2.2. There are $n!$ choices of $T$. Out of them we choose the map $T_0$, which minimises the following sum
$$ \sum_{i=1}^n \| X_i - T(X_i)\| .$$
Suppose now the vectors $(r_k)_{k=1}^d$ are formed as
\begin{equation}\label{OTr}
r_{k,i} = \frac{1}{\sqrt n} r_k(T_0 (X_i)), \;  \; k=1,\dots, d.
\end{equation}
Here $(r_k(\cdot))_{k=1}^d$ is a system of orthonormal functions on $L_2[0,1]^p$.  With this choice of $(r_k)_{k=1}^d$, define residuals $\hat e$ again as \eqref{param-e}. Justification of the use of the operator $T_0$ partly comes from equality
\begin{equation}\label{OTunif}
G_n(x) = \frac{1}{n}\sum_{i=1}^n {\mathbb{I}}_{(T_0(X_i)\leq x)} =\frac{1}{n}\sum_{i=1}^n {\mathbb{I}}_{(\xi_i\leq x)} ,
\end{equation}
which shows that $G_n$ will converge to the uniform distribution function on $[0,1]^p$. As a corollary of \eqref{OTr} and \eqref{OTunif}, the behaviour of statistics, which are invariant under permutations, is governed by $G_n$ and not by $F_n$. For example
\begin{equation}\label{OTrI}
\frac{1}{n}\sum_{i=1}^n r_{k}(T_0(X_i)) {\mathbb I}_{(T_0(X_i)\leq x)} = \int_{z\leq x} r_k(z) dG_n(z).
\end{equation}

Using $T_0$ we can transform the process $\hat w_{n,e}$ of Proposition \ref{prop2}, $(ii)$, as follows:
\begin{equation}\label{OT}
T_0^* \hat w_{n,e}(x) = \frac{1}{\sqrt {n}}\sum_{i=1}^n \hat e_i {\mathbb{I}}_{(T_0(X_i)\leq x)},
\end{equation}
where the construction of $\hat e$ incorporates, as we said, $T_0(X_i)$-s. The following comment is intended as further justification of the use of $T_0$. It is not necessary to use minimiser $T_0$ to produce the version of regression empirical process with standard covariance operator -- any $T$ will achieve this. However, in the case when the null hypothesis \eqref{gl} is not correct, expected values of residuals $\hat e$ are not zero, but will be, for each contiguous converging alternatives, close to some function, say, $h$, specific to the alternative (see, e.g., \cite{KhmKou04}, sect. 1, or \cite{HajSid67}). It will be desirable that the shift of transformed process $T_0^* \hat w_{n,e}$ preserves the main pattern present in the shift function $h$. For this, it is necessary that the transformation of $\hat w_{n,e}$ be smooth. One can say that the $T$ should minimise the sum
$$ \sum_{i=1}^n | h(X_i) - h(T(X_i)) | .$$
However, very wide class of alternatives, and therefore, of functions $h$ is apriori possible.
Therefore, the choice of $T$ should not be hinged on a particular $h$ but should be as ``smooth" map of  $(X_i)_{i=1}^n$ into $(\xi_i)_{i=1}^n$ as possible. This motivates the choice of $T_0$.

We formulate the next proposition  for readers' convenience. It does not require a new proof, and we will give only short comments at the end of it. 
\begin{prop}\label{prop3} Suppose the regression function $m_\theta(x)$ is regular, in the same sense as in Proposition \ref{prop2}.
Suppose the orthonormal functions $r_k(\cdot), k=1,\dots,d,$ are continuous and bounded on $[0,1]^p$.Then

(i) for the covariance matrix of the residuals $\hat e$ the following is true: 
\begin{equation*}
E\hat e \hat e^T \sim I - \sum_{j=1}^d  r_k r_k^T, \; n\to\infty,
\end{equation*}
where $r_k$ are realised according to \eqref{OTr};

(ii) for the empirical regression process, based on residuals $\hat e$ of \eqref{param-e},
$$T_0^* \hat w_{n,e}(x) = \frac{1}{\sqrt {n}}\sum_{i=1}^n \hat e_i {\mathbb{I}}_{(T_0(X_i)\leq x)},$$
the following convergence of the covariance function is true:
$$ET^*_0\hat w_{n,e}(x) T^*_0\hat w_{n,e}(y) \to G(\min(x,y)) - \sum_{k=1}^d Q_k(x) Q_k(y) , \text {as} \; n\to \infty ,$$
where $Q_k(x)=\int_{z\leq x} r_k(z) dz$;\\
moreover, 

(iii) the process $T_0^*\hat w_{n,e}$ converges in distribution to projection of standard Brownian motion on $[0,1]^p$ orthogonal to functions $r_k(\cdot), k=1,\dots, d$.
\end{prop}

Given two orthonormal systems of $n$-vectors $(\mu_{\theta k})_{k=1}^d$ and $(r_k)_{k=1}^d$ the operator $K_d$ will rotate one system into another, regardless of how these systems have been constructed. Therefore \eqref{strong_map} is also true for $p$-dimensional time, and this implies (i).

To see that (iii) is true denote ${\mathbb{I}}_{T_0, x}$ the vector with coordinates ${\mathbb{I}}_{(T_0(X_i)\leq x)}$. Now we use \eqref{strong_map} to write the process $T_0^*\hat w_{n,e}$ in the form
$$
T_0^* \hat w_{n,e}(x) \sim \frac{1}{\sqrt n} \langle\mathbb{I}_{T_0, x},\epsilon - \sum_{k=1}^d r_k \langle r_k, \epsilon\rangle \rangle,  
$$
and then use the representation of $(r_k)_{k=1}^d$ through functions $(r_k(\cdot))_{k=1}^d$:
$$\frac{1}{\sqrt n} \langle\mathbb{I}_{T_0, x},\epsilon \rangle = \frac{1}{\sqrt n} \sum_{i=1}^n {\mathbb{I}}_{(T_0(X_i)\leq x)} \epsilon_i = T^*_0 w_n(x)$$
\begin{align*}\frac{1}{\sqrt n} \langle \mathbb{I}_{T_0,x},r_k \rangle  \langle r_k, \epsilon\rangle &= \frac{1}{n}\sum_{i=1}^n {\mathbb{I}}_{(T_0(X_i)\leq x)} r_k(T_0(X_i)) \frac{1}{\sqrt n} \sum_{i=1}^n r(T_0(X_i)) \epsilon_i\\
&=\int_0^x r(z) dG_n(z) \int r_k(z) T^*_0 w_n(dz)
\end{align*}
This altogether leads to
$$T_0^*\hat w_{n,e}(x) \sim T^*_0 w_n(x) - \sum_{k=1}^d \int_0^x r_k(z) dG_n(z) \int r_k(z) T^*_0 w_n(dz).$$
The process $T^*_0 w_n$ obviously converges to $G$-Brownian motion (that is, standard Brownian motion) on $[0,1]^p$, while $T^*_0 \hat w_{n,e}$ differs from it by the term which involves only finitely many linear functionals from it. 

We formulated (ii) for the sake of some symmetry of presentation. To see that (ii) is true, one can follow the proof of (ii) in Proposition \ref{prop2} using \eqref{OTunif} in place of  $\mathbb{I}_{x}^T \mathbb{I}_{y}$,
$$G_n(\min(x,y)) = \frac{1}{n} \sum_{i=1}^n  {\mathbb{I}}_{(T_0(X_i)\leq x)} {\mathbb{I}}_{(T_0(X_i)\leq y)} =\frac{1}{n}\sum_{i=1}^n  {\mathbb{I}}_{(T_0(X_i)\leq \min(x,y))} $$
and using \eqref{OTrI} in place of $\frac{1}{\sqrt n}\mathbb{I}_{x}^T  r_k $. y. On the other hand, it also follows from (iii).\\

\begin{figure}[h!]
\includegraphics[width=14.5cm, height=15.6cm]
{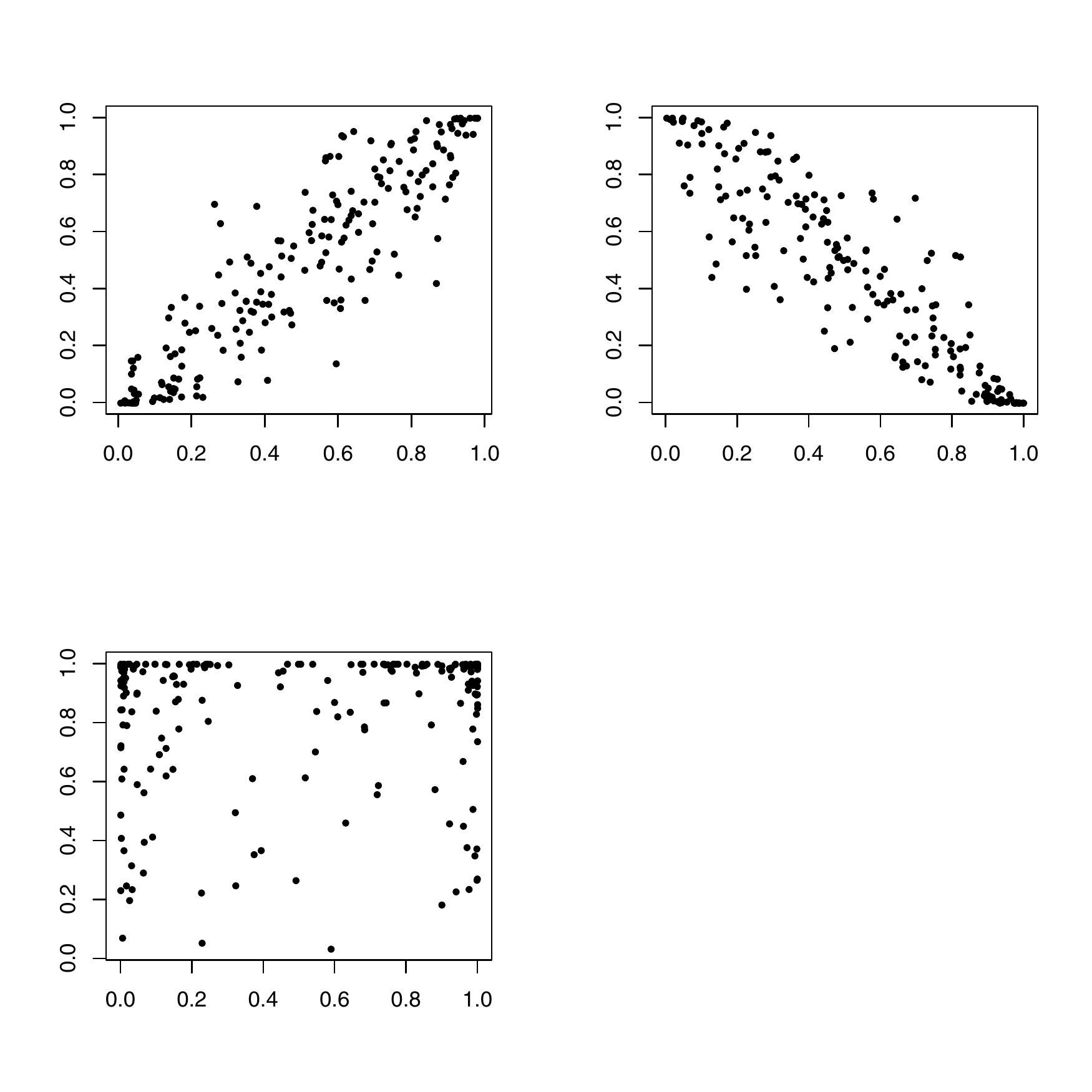}
\caption{In the three scatterplots the covariates $(X_i)_{i=1}^n$ are generated as 2-dimensional iid random variables, but in the first row coordinates of each $X_i$ are not independent: they are $X_{i1}\sim {\cal U}[0,1], X_{i2}\sim B(8(1-X_{i1}), 8 X_{i1})$ on the left scatter-plot, and $X_{i1}\sim {\cal U}[0,1], X_{i2}\sim B(8X_{i1}, 8 (1-X_{i1}))$ on the right one. On the third scatterplot the coordinates are independent, but have different $B$-distributions: $X_{i1}\sim B(0.35,0.35)$ and $X_{i2}\sim B(0.2,0.2)$}
\label{Fig:Scatters}
\end{figure}

In order for the optimal transport method to work one does not need continuity of the limiting distribution $F$. One only needs $n$ distinct points $(X_i)_{i=1}^n$ in the unit square. It is also not necessary that $(\xi_i)_{i=1}^n$ be generated as random variables -- they can be strategically placed to form a uniformly spread net. On the other hand, to find a minimiser $T_0$ can be computationally costly, more so than the estimation of density based on $F_n$, if one employs the transformation described in \cite{Khm16}. More detailed comparison of the two methods are the subject of the paper \cite{Ban19}. 

\begin{center}
\begin{figure}[h!]
\includegraphics[width=13cm, height=6.5cm]{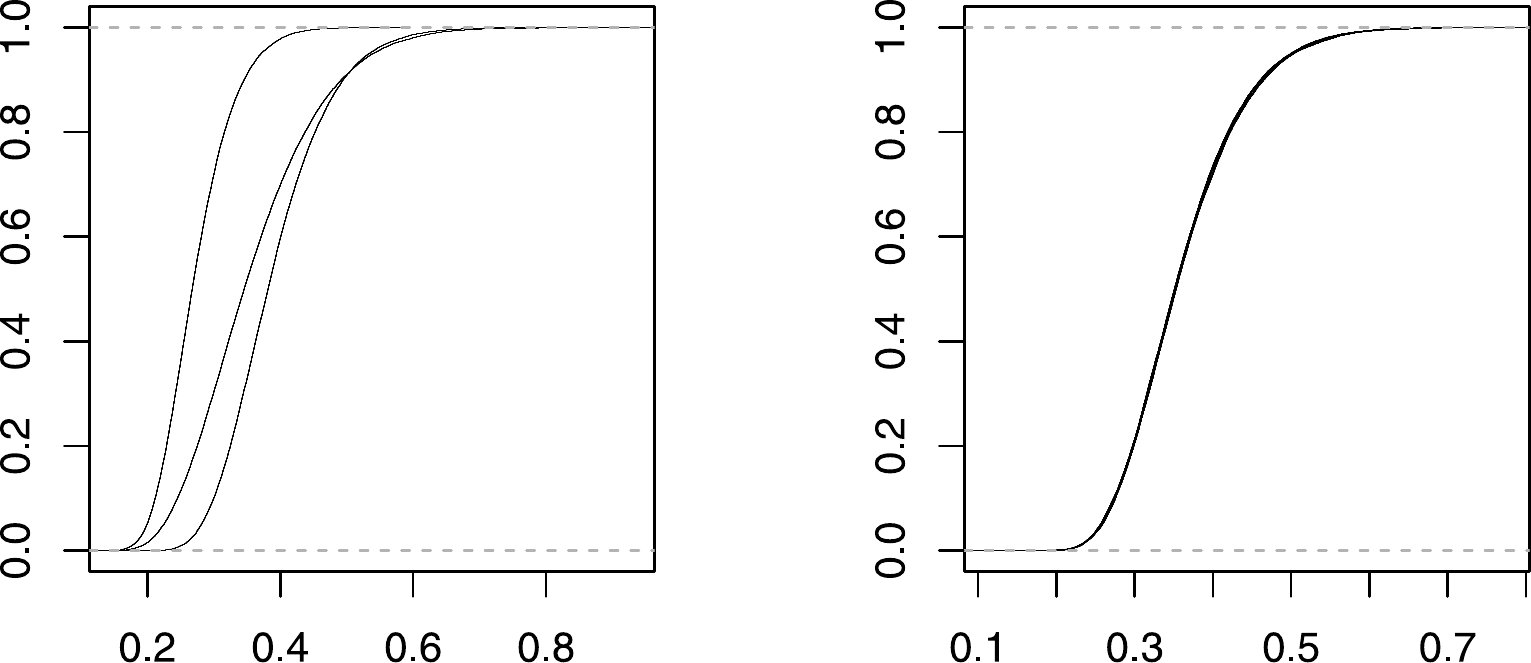}
\caption{On the left panel we show three simulated distribution functions of statistic $D_n^+=\max_x \hat w_n(x)$  for $X_i$-s distributed as on the three scatterplots shown above. These distribution functions are indeed different. On the right panel there are also three graphs of distribution functions of the statistic from the transformed process $D_{n,e}^+=\max_{x}T_0^* \hat w_{n,e}(x)$ for the same three scatterplots. Sample size in all cases was $n=200$. Visually the graphs are indistinguishable.}
\label{Fig:KS-multi}
\end{figure}
\end{center}

\section{On power considerations}

We do not advocate in this paper any particular test. Any test based on a functional from the transformed empirical process $T_0^* \hat w_{n,e}(x)$ is asymptotically distribution free, and which particular functional will be chosen remains in discretion of a user.

On the other hand, distribution free-ness can not be the only requirement on a statistic or an underlying empirical process, because trivial and useless choices are possible. The version of regression empirical process constructed in this paper satisfies two requirements, not one: a) under the null hypothesis its limit distribution does not depend on parametric family of regression functions or the true value of the parameter, and b) for any sequence of alternative regression functions $b_n$, converging to $m_\theta$ at some $\theta$ from the (functional) direction $\phi$,
$$b_n(x)=m_\theta(x)+\frac{1}{\sqrt{n}} \phi_n(x), \quad \int [\phi_n(x) - \phi(x)]^2 dF(x) \to 0,$$
the  statistic of locally most powerful test for testing against the sequence $b_n$ is a functional of the transformed regression empirical process. So, it is asymptotically distribution free and sensitive to all local alternatives at the same time.

Note that the regression empirical process $\hat w_n$ does have the property b) (cf.,e.g., \cite{KhmKou04})
and the process $T_0^* \hat w_{n,e}(x)$ being its ``smooth" one-to-one transformation, also has this property. This also implies that test statistic based on $\hat w_n$  can be viewed as a statistic based on  $T_0^* \hat w_{n,e}(x)$, and vice versa. Therefore, at the first glance natural question on power behaviour of the ``same test" from the two processes is only a question of comparing two different tests from the same empirical process. This is the case, for example, for two Kolmogorov-Smirnov statistics
$$D_{n} = \max_x |\hat w_n(x)| \quad \text{and}\quad D_{n,e}=\max_x |T_0^* \hat w_{n,e}(x)| ,$$
or the second maximum taken from $\hat w_{n,e}(x)$ if covariates are one-dimensional. For a reader with some experience in goodness of fit theory it will be clear that both tests are admissible, therefore neither dominates the other in statistical power.

\begin{center}
\begin{figure}[h!]
\includegraphics[width=12.5cm, height=7.5cm]{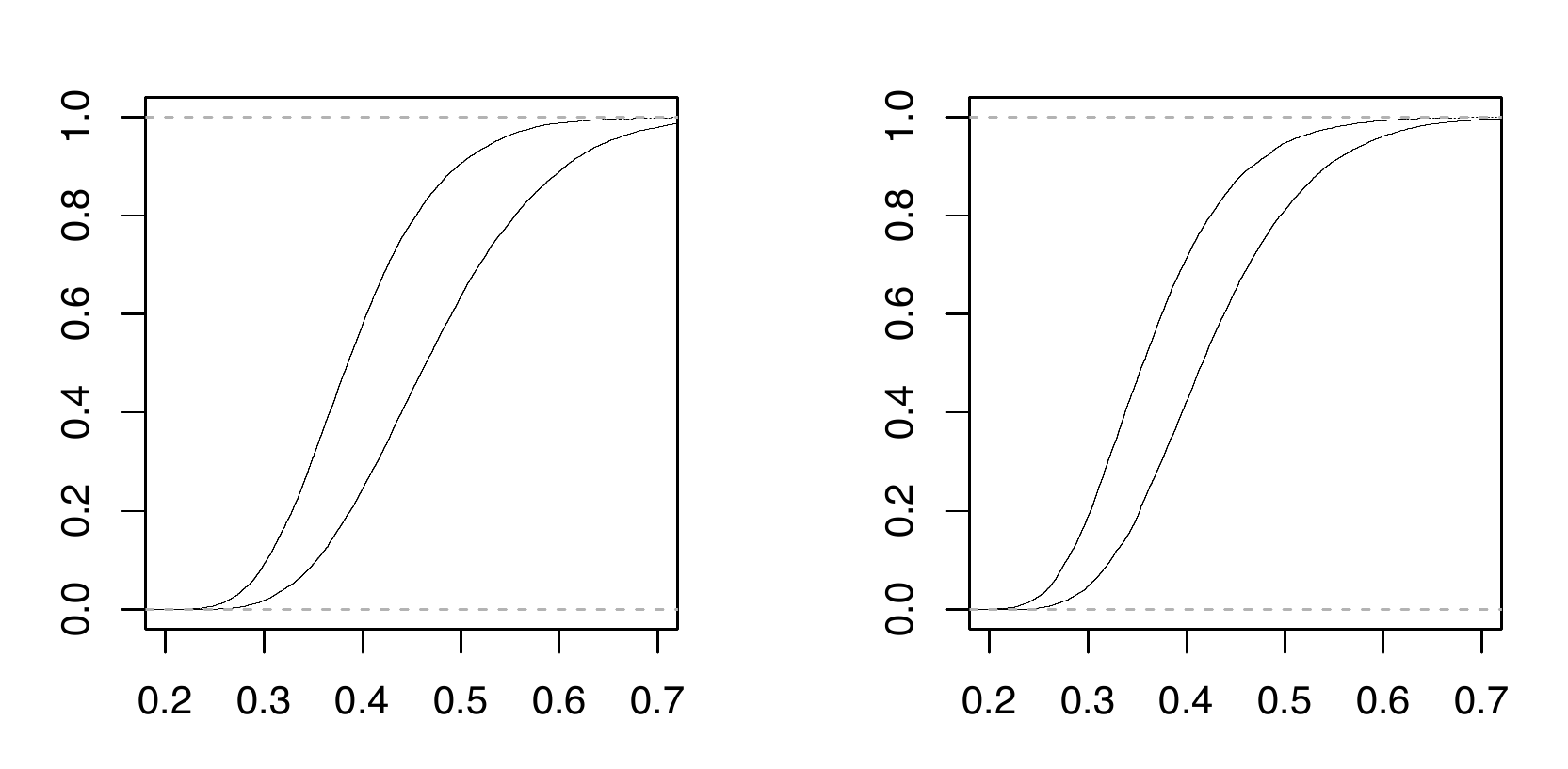}
\caption{Here $\psi(x)=x_2^3$ and sample size $n=200$. Although the uniform distance, and therefore the distance in total variation, between the two distributions on both panels are very similar, the overall impression well may be that $D_{n}$, the KS statistic from unmodified regression process (left panel), reacts on the alternative somewhat better than $D_{n,e}$.}
\label{Fig:KS-power1}
\end{figure}
\end{center}

\begin{center}
\begin{figure}[h!]
\includegraphics[width=12.5cm, height=8.5cm]{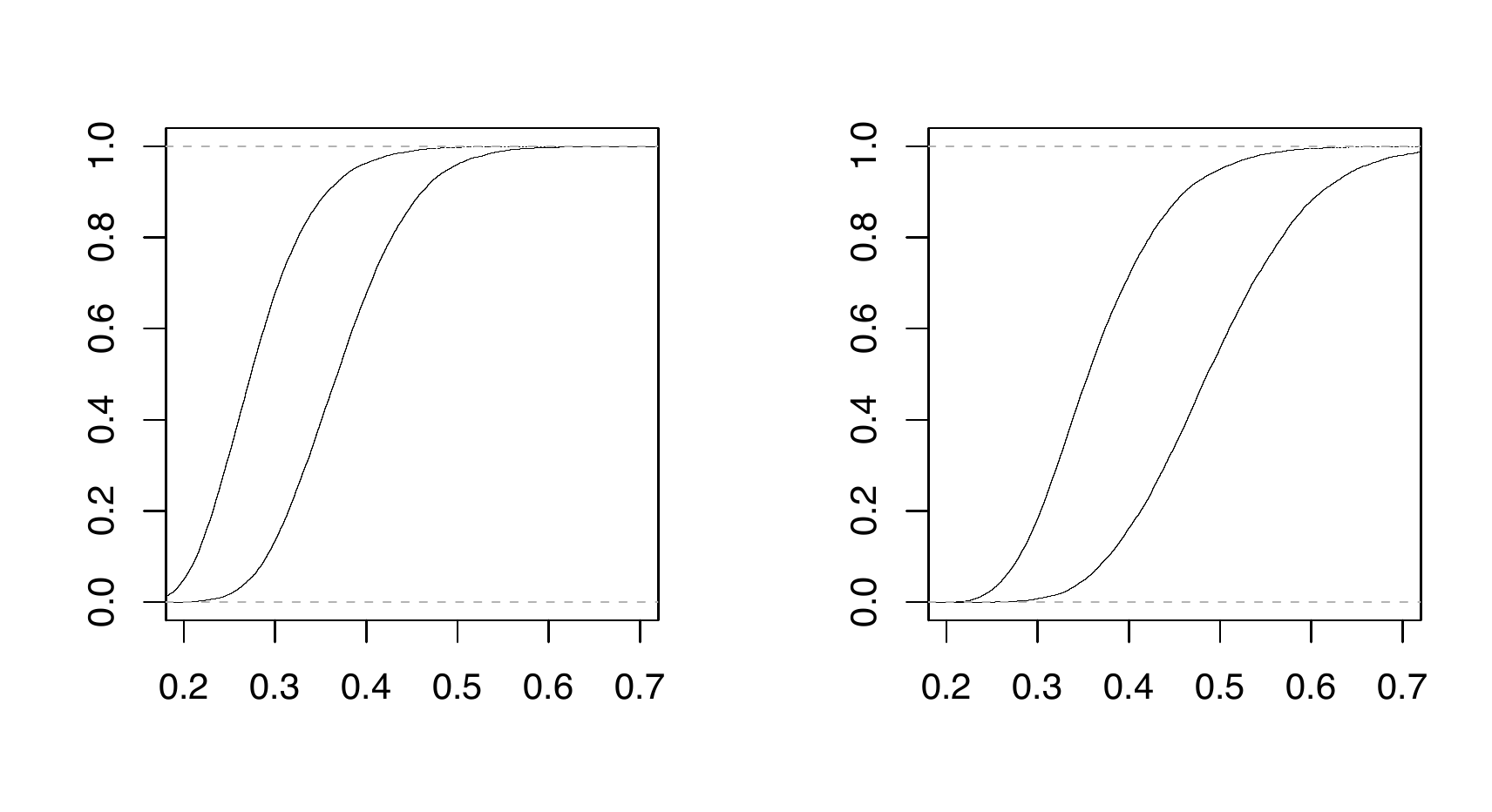}
\caption{Here $\psi (x)=\sin(\pi x_2/2)$. and sample size $n=200$. Although the uniform distance between the two distributions on both panels is still similar, the overall impression is that $D_{n,e}$, the KS statistic from the transformed regression empirical process (right panel), reacts on the alternative better than $D_{n}$.}
\label{Fig:KS-power2}
\end{figure}
\end{center}

Here is an illustration of this point in two more figures. The left panel in Figure \ref{Fig:KS-power1} shows distribution functions of statistic $D_{n} = \max_x |\hat w_n(x)|$ under the null model, with two-dimensional covariates and with $$m_\theta(X_i)=\theta_0 + \theta_{10} (X_{1i} - \widebar{X_{1n}}) + \theta_{01} (X_{2i} - \widebar{X_{2n}}) + \theta_{11} (X_{1i}X_{2i} - \widebar{X_1 X_2})$$
and under alternative $m_\theta(x)+ x_2^2$, while the right panel shows the distributions of statistic  $D_{n,e}=\max_x |T_0^* \hat w_{n,e}(x)|$ in the same situation. Figure \ref{Fig:KS-power2} shows the situation under the same model, but now with $b_n(x) = m_\theta (x) + \sin (\pi x_2/2)$.

To complement short discussion in the previous section on why we need to use the optimal transport $T_0$ note the following: as we remarked, choice of the optimal transport map will transform the shape of the bias term $\psi$ in consistent way, but  one needs to be sure that this consistency is preserved as $n\to\infty$. This latter is true, however, as it follows, e.g., from \cite{CueMatTue97}           Theorem 3.2.

 \section{Acknowledgment} 
For the results, shown on Fig \ref{Fig:Scatters} -- Fig \ref{Fig:KS-power2} and many more experiments, not included here, the author is grateful to his student Richard White.

\end{document}